\documentclass[preprintnumbers,amsmath,amssymb,floatfix,12pt,prd,superscriptaddress,nofootinbib]{revtex4}
\usepackage{graphicx}
\usepackage{epsfig}
\usepackage{bm}
\usepackage{amsfonts}

\def\ben{\begin{equation}}
\def\een{\end{equation}}

 \def\bd{\begin{document}} \def\ed{\end{document}}
\def\ds{\documentstyle} \let\fr=\frac \let\bl=\bigl \let\br=\bigr
\let\Br=\Bigr \let\Bl=\Bigl
\let\bm=\bibitem
\let\na=\nabla
\let\pa=\partial \let\ov=\overline
\newcommand{\be}{\begin{equation}}
\newcommand{\ee}{\end{equation}}
\def\ba{\begin{array}}
\def\ea{\end{array}}
\def\ft#1#2{{\textstyle{\frac{\scriptstyle #1}{\scriptstyle #2} } }}
\def\fft#1#2{{\frac{#1}{#2}}}
\def\del{\partial}
\def\vp{\varphi}
\def\sst#1{{\scriptscriptstyle #1}}
\def\oneone{\rlap 1\mkern4mu{\rm l}}
\def\td{\tilde}
\def\wtd{\widetilde}
\def\ie{{\it i.e.\ }}
\def\dalemb#1#2{{\vbox{\hrule height .#2pt
        \hbox{\vrule width.#2pt height#1pt \kern#1pt
                \vrule width.#2pt}
        \hrule height.#2pt}}}
\def\square{\mathord{\dalemb{6.8}{7}\hbox{\hskip1pt}}}
\newcommand{\ho}[1]{$\, ^{#1}$}
\newcommand{\hoch}[1]{$\, ^{#1}$}
\newcommand{\bea}{\setlength\arraycolsep{2pt} \begin{eqnarray}}
\newcommand{\eea}{\end{eqnarray}}
\newcommand{\ra}{\rightarrow}
\newcommand{\lra}{\longrightarrow}
\newcommand{\Lra}{\Leftrightarrow}
\newcommand{\bp}{\tilde \beta^\prime}
\newcommand{\tr}{{\rm tr} }
\newcommand{\Tr}{{\rm Tr} }
\def\0{{\sst{(0)}}}
\def\1{{\sst{(1)}}}
\def\2{{\sst{(2)}}}
\def\3{{\sst{(3)}}}
\def\4{{\sst{(4)}}}
\def\5{{\sst{(5)}}}
\def\6{{\sst{(6)}}}
\def\7{{\sst{(7)}}}
\def\8{{\sst{(8)}}}
\def\m{{\sst{(m)}}}
\def\n{{\sst{(n)}}}
\def\cA{{{\cal A}}}
\def\cB{{{\cal B}}}
\def\cF{{{\cal F}}}
\def\cG{{{\cal G}}}
\def\cH{{{\cal H}}}
\def\tV{\widetilde V}
\def\tW{\widetilde W}
\def\tH{\widetilde H}
\def\tE{\widetilde E}
\def\tF{\widetilde F}
\def\tA{\widetilde A}
\def\im{{{\rm i}}}
\def\tY{{{\wtd Y}}}
\def\ep{{\epsilon}}
\def\vep{{\varepsilon}}
\def\bD{{{\bar D}}}
\def\R{{{\mathbb R}}}
\def\C{{{\mathbb C}}}
\def\H{{{\mathbb H}}}
\def\CP{{{\mathbb C}{\mathbb P}}}
\def\RP{{{\mathbb R}{\mathbb P}}}
\def\Z{{{\mathbb Z}}}
\def\bA{{{\mathbb A}}}
\def\bB{{{\mathbb B}}}
\def\bC{{{\mathbb C}}}
\def\bD{{{\mathbb D}}}
\def\bE{{{\mathbb E}}}
\def\bZ{{{\mathbb Z}}}
\def\Re{{{\frak{Re}}}}
\def\Im{{{\frak{Im}}}}
\def\cosec{{\,\hbox{cosec}\,}}
\def\Gm{{\Gamma_{\!\! -}}}
\def\Gp{{\Gamma_{\!\! +}}}
\def\stan{{standard }}
\def\nonstan{{supernumerary }}
\def\p{{\partial}}
\def\kdel#1{{\fft{\del}{\del#1}}}

\def\bog{{Bogomolny }}
\def\om{{\omega}}

\newcommand{\nnr}{\nonumber \\}
\newcommand{\pd}{\partial}
\newcommand{\ud}{\textrm{d}}
\newcommand{\dTH}{T^{\prime \, 0}_\textrm{H}}
\newcommand{\dOi}{\Omega^{\prime \, 0}_i}
\newcommand{\bx}{{\bf x}}
\begin{document}
\title{Time varying gravitational constant  $G$ via the entropic force}
\author{\textbf{M. R. Setare}} \email{rezakord@ipm.ir}
\affiliation{Department of Science, Payame Noor University. Bijar.
Iran}
\author{\textbf{D. Momeni}}
\email{d.momeni@yahoo.com} \affiliation{Department of Physics ,
Faculty of Science,
  Tarbiat Moallem University, Tehran , Iran}

\begin{abstract}
\vspace*{1.5cm} \centerline{\bf Abstract} \vspace*{0.5cm} If the
uncertainty principle applies to the Verlinde entropic idea, it
leads to a new term in the  Newton's second law of mechanics in the
Planck's scale. This curious velocity dependence term inspires a
frictional feature of the gravity. In this short letter we address
that this new term modifies the effective mass and the Newtonian
constant as the time dependence quantities. Thus we must have a
running on the value of the effective mass on the particle mass $m$
near the holographic screen and the $G$. This result has a nigh
relation with the Dirac hypothesis about the large numbers
hypothesis (L.N.H.) [1]. We propose that the corrected entropic
terms via Verlinde idea can be brought as a holographic evidence for
the authenticity of the Dirac idea.
\end{abstract}
 \pacs{04.70.Bw, 11.25.Tq, 74.20.-z}
 \keywords{Gauge-gravity correspondence, Models of Quantum Gravity}
\newpage
\maketitle
\section{Introduction}
Recently Verlinde presented a new approach \cite{ver} to the nature
of the gravitational interaction by a Holographic picture. To
illuminate the relation between gravity and thermodynamics, there
are some momentous works back to the Padmanabhan which was appeared
before Verlinde paper\cite{pd1,pd2}. Verlinde's simple formalism
swindles every body to interpret the gravity as an interaction
between a screen with coding amount of the information and a test
particle. It was shown that this form of the the interaction  causes
a slight change in the entropy. There are many publications on
entropic formalism, inspired directly from the work of the
Verlinde\cite{ref}. The information that was saved on the screen
must be modified by some further quantum considerations as the
Generalized Uncertainty Principle (GUP) \cite{gup} idea for adding
some corrections to the Newtonian gravity. One of the most important
corrections is the effect of the minimum Plank's length and the
modification of the entropic force. As it was shown in \cite{ghosh}
this correction which comes from the GUP, is induced as the friction
term to the force law, and if we accept that this friction causes
motion under a dissipative force then the question is why this term
is necessary in the description of the gravity?. Answering to this
question is the main goal of us in this short letter. We explicitly
show this new dissipative force via the GUP, forces the running to
the Newtonian constant $G$ and in a weaker statement the descend of
it for future times. The test particle mass also involves similar
phenomenon which is the running of the effective mass of the test
particle.

\section{Planck Scale corrections to the gravitational law}
As it had been shown in Refrence \cite{vancea} if one introduces
Planck scale's minimum length induced directly from the GUP (which
is a fundamental counting the cells in the holographic screen ) the
quantum makes correct the Newton's second law of motion
\begin{eqnarray}
\overrightarrow{F}=m\overrightarrow{a}\nu-\frac{\overrightarrow{p}}{\tau}
\end{eqnarray}
where
\begin{eqnarray}
\nu\equiv1+\frac{1}{2}(\frac{\lambda}{\delta x})^{2}\frac{p}{mc},
\hspace{0.5cm} \tau^{-1}\equiv\frac{c\lambda}{2(\delta x)^{2}},
\end{eqnarray}
here $\lambda=\frac{\hbar}{mc}$ is the particle's Compton length,
$\delta x$ is the displacement of the particle near the screen, and
$\overrightarrow{p}$ is the momentum of a particle. For the Kepler
i.e. the motion of a particle in a centrifugal force's field,
 the differential equation (DE) (1) is in the form of the second kind of Abel's
equation, class A and as we know that there is no general solution
for this DE yet \cite{de}. Only if we neglect the gravitational
force and solve the ODE (1) for a free particle we can obtain the
following exact result for momentum $p$ in terms of the Lambert
function\footnote{ Consideration of Lambert W function can be traced
back to J. Lambert around 1758, and later, it was considered by L.
Euler but it was recently established as a special function of
mathematics on its own[12]. The Lambert W function is defined to be
the function satisfying $W[z]e^{W[z]} = z $ It is a multivalued
function defined in general for z complex and assuming values W[z]
complex. If z is real and $z < -1/e$, then W[z] is multivalued
complex. If z is real and $-1/e < z < 0$, there are two possible
real values of W[z]. The one real value of W[z] is the branch
satisfying $ W[z]\leq -1$, denoted by $W_{0}[z]$, and it is called
the principal branch of the W function. The other branch is $
W[z]\leq -1$ and it is denoted by $W_{-1}[z]$. If z is real and $z
\geq 0$, there is a single real value for W[z] which also belongs to
the principal branch $W_{0}[z]$. Special values of the principal
branch of the Lambert W function are $W_{0}[0] = 0$and $W_{0}[-1/e]
= -1$. The Taylor series of $W_{0}[0]$ about $z = 0$ can be found
using the Lagrange inversion theorem and is given by [13]
\begin{eqnarray}\nonumber
W[z]=-z\sum_{n=0}^{\infty}\frac{(-z)^{n}(n+1)^{n}}{n!}
\end{eqnarray}
The ratio test establishes that this series converges if $\mid
z\mid<1/e$.}

\begin{equation}\nonumber
p=e^{-\frac{\tau W(t)-t-c_1}{\tau}}
\end{equation}
Where $W(t)=LambertW((\frac{\lambda}{\delta
x})^2\frac{e^{\frac{t+c_1}{\tau}}}{2mc}) $. The Fig.1 shows the
behavior of the rescaled momentum p for some values of the parameter
$\tau$.
\begin{figure}
\centering
 \includegraphics[width=10cm,angle=0] {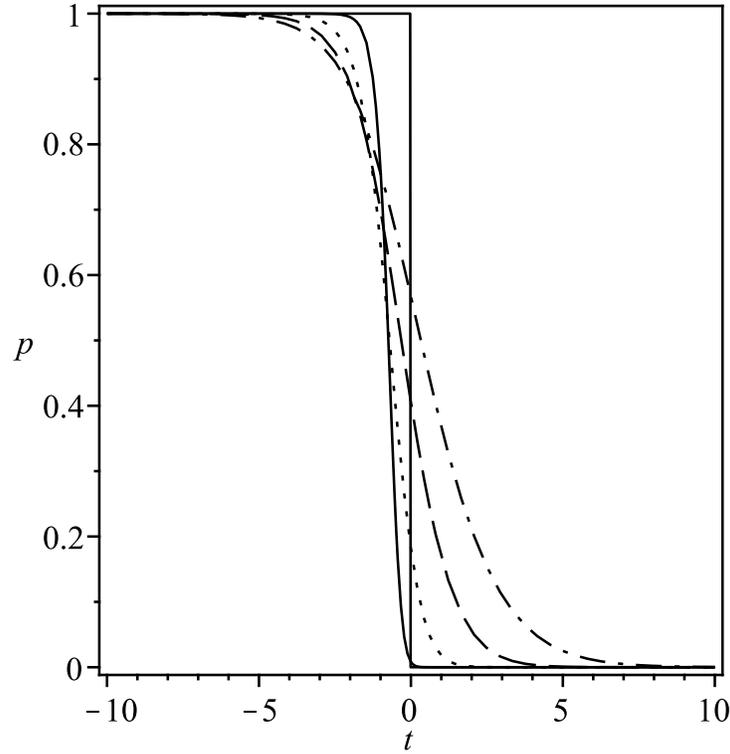}
  \caption{Variation of the $ p$ for a sample of the parameter $\tau$. The value of $\tau$ from down to top
is equal to $1,0.75,0.5,0.25,10^{-4}$.}
 \label{1.eps}
\end{figure}

 For small momentum by
transferring $p$ in the regime of non relativistic test particle
near holographic screen $\frac{p}{mc}<<1$ we can neglect from the
second term of the $\nu$  in (2) and take it as unity i.e.
$\nu\simeq 1$. We mention here that a relativistic gravitational
field can not be considered and as we know that which further
detailed that requires more general relativity theory. In the
Verlinde original proposal for gravity he neglected the curve space
effects at the first step. Thus we set $\nu=1$.

\begin{figure}
\centering
 \includegraphics[width=10cm,angle=0] {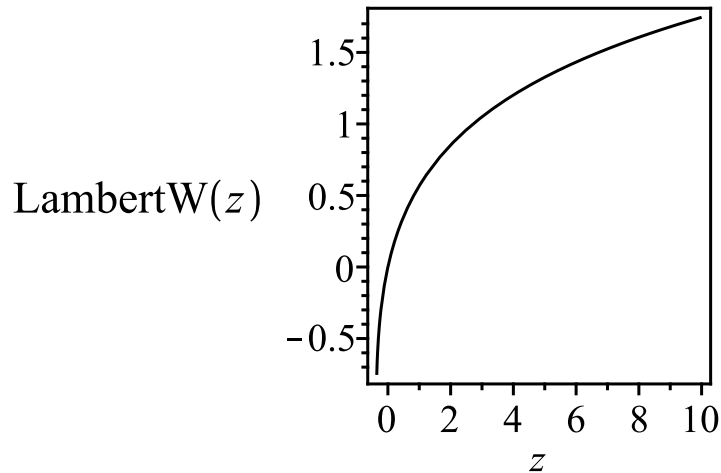}
  \caption{Variation of the $LambertW(z) $ .}
 \label{2.eps}
\end{figure}
\section{The Hamiltonian of a test particle}
The simple time dependence Hamiltonian for the system (1) in a
conservative central force field
\begin{eqnarray}
 V(r)=-\frac{K}{r}
\end{eqnarray}
where   $K=GMm$   is read from the comparing (1) with a one
dimensional dissipative simple harmonic oscillator in classical
mechanics \cite{goldestein}. We find the Hamiltonian describing the
test particle of mass m moving under the influence of the central
potential to be
\begin{eqnarray}
H=\frac{\overrightarrow{p}.\overrightarrow{p}}{2m}e^{\frac{t}{\tau}}+e^{-\frac{t}{\tau}}V(r)
\end{eqnarray}
our Hamiltonian is rotational invariant exactly as the $\kappa$
 spacetime like in a noncommutative version of the Kepler's problem
 \cite{nc}. Also we note that the modification for the
kinetic term can be captured by a suitable definition of the mass as
an effective mass $m_{eff}$
\begin{eqnarray}
m_{eff}=m e^{\frac{-t}{\tau}}
\end{eqnarray}
Here, we also note that replacing  m  with $m_{eff}$ in Lagrangian
does not affect the rotational invariance of the system described by
above Hamiltonian. Comparing (4) with (5) testifies that the
superficial constant of the $k$ must have the following running:
\begin{eqnarray}
K=GMme^{\frac{-t}{\tau}}
\end{eqnarray}
Note that (6) recovers the classical results in limit
$\hbar\rightarrow0, \tau\rightarrow\infty$ but queer in a curious
running scheme. From 5 we conclude that in the entropic scenario,
the screen does not sense the usual classical mass $m$ but it senses
an effective time dependence mass which it  decays with time  in GUP
regime. This feature shows that in general, there are more
ambiguities for applying the Verlinde idea based on the usual
equilibrium statistical mechanics. In the next sections we show that
the modification of the usual gravitational constant is a good
evidence for proposing a new Dirac large numbers hypothesis. But
before doing it, we review Dirac's idea about the existence of
dimensionless large numbers in nature.
\section{Dirac Large Numbers Hypothesis and  entropic running of the $G$}
In this section first we review the Dirac's theory and then we will
show some similarities of our result with Dirac's hypothesis.
\subsection{Dirac Large Numbers Hypothesis}
There are three dimensionless numbers in the nature which can be
constructed from the atomic and cosmological datas:\\
1-The ratio of the electric to the gravitational force between an
electron and a proton $7\times10^{39}$\\
2-The age of the Universe t, expressed in terms of a unit of time
provided by atomic constants $\frac{e^{2}}{m_{e}c^{3}}$\\
and finally \\
3-\emph{ The mass of that part of the Universe that is receding from
us with a velocity $v<c/2$
expressed in units of the proton mass of the order $10^{78}$}\\
Dirac large number
hypothesis in it's orthodoxies form Germaned by himself says that\\
"..\emph{these numbers are related by equations in which the
coefficients are close to unity}". Since the number in (2) varies
with the age of the Universe, the L.N.H. requires that the other
numbers must also vary, namely
\begin{eqnarray}
\frac{e^{2}}{Gm_{e}m_{p}}\propto t
\end{eqnarray}
or
\begin{eqnarray}
N\propto t^{2}\\
G\propto t^{-1}
\end{eqnarray}
There are two interpretations for the above relation both discussed
by Dirac and the only one which was acceptable by himself as
\emph{One can reconcile the relation (9) with conservation of mass
by assuming that the velocity of recession of a galaxy is
continually decreasing, so that more and more galaxies are
continually appearing
with velocity of recession $<c/2$. }\\
This is the picture which was adopted in his first paper on the
subject \cite{1938}. There are a serious problem between (9) and GR:
the Einstein's theory requires G to be constant. As was noted by
Dirac's theory this inconsistency might be solved if\emph{".. we
assume that the Einstein theory is valid in a different system of
units from those provided by the atomic constants."}
\subsection{Entropic corrections as a new L.N.H. }
First we write (6) as the following form:
\begin{eqnarray}
G=G_{N} e^{\frac{-t}{\tau}}
\end{eqnarray}
Where the $G_{N}$ is the usual Newtonian gravitational constant. The
quantum Planck's corrections from GUP are significance only at
Planck's time scales for example in GUT epoch. Thus essentially the
time coordinate $t$ is very small in comparison with the new scale
$\tau$. We can expand the exponential term in (10) in powers of the
$t/\tau$:
\begin{eqnarray}
G=G_{N} (1-\frac{t}{\tau}+\frac{t^{2}}{2\tau^{2}}+..)
\end{eqnarray}
Thus for enough small times comparable with $\tau$ the linear term
 dominates,thus the new L.N.h. is
\begin{eqnarray}
N\propto (1- (\frac{t}{\tau})^2)\\
G\propto(1- \frac{t}{\tau})
\end{eqnarray}
\section{ Comparing the model with observational constraints}
From observational view, as it was shown \cite{pitjeva}, there is a
secular decrease of the gravitational constant $G(t)$ . This time
rate of change is of order $\frac{\dot{G}}{G}=(-5. 9\pm 4.4)\times
10^{-14} yr^{-1}$ which it has been obtained from planetary data
analysis. It is convient that we compare (10) with observational
bounds. Remembering to mind that, the time scale of our model is
$\tau=\frac{2(\delta x)^2}{c\lambda}$. It's physically reasonable ,
we compare this time scale in regime that the particle's replacement
$\delta x$ be of order of the Compton's wave length of the test
particle. This means that $\delta x\sim O(\lambda)$. For example ,
we take $\delta x=\lambda$ , then we lead to
$\tau=\frac{2\lambda}{c}$. It is possible to show that in this case
$\frac{\dot{G}}{G}=-\frac{2mc^2}{h}$. We take the test particle as
an electron with mass $m=0.511 Mev/c^2$, then $\frac{\dot{G}}{G}=O(
10^{-14} yr^{-1})$. Thus, our theoretical result is compatible with
the observational reports, if we take the element of the particle's
displacement of order the Compton wave length of the electron, as a
test particle.

\section{Summary}
Briefly we show that the new corrections from GUP were imposed on
the entropic nature of the gravity,by possessing a running value for
the $G$. This simple derivation shows that there is a ticklish
relation between the entropic corrections and the GUP and also on
which happened in a $\kappa$ space analysis of the Kepler's problem
in non commutative spacetimes. It is a good idea to investigate the
relation between non commutative spacetimes and the entropic origin
of the gravity. Also there is a L.N.h. like relation for this
theory. Thus there must exist a tight relation between Holographic
description of the Gravity, non commutativity, L.N.h. Thus any
unification of the Gravity and the Quantum physics will be hanker
between these features.
\section{Acknowledgment}
The authors would like to thank E. Verlinde  for a private
communication and pointing some errors in the first version of this
work. The authors would like to thank anonymous referees  for
helpful comments and suggestions.  D. Momeni also acknowledges the
support by Tarbiat Moallem University, Tehran , Iran.

\end{document}